# NON-EQUILIBRIUM THERMODYNAMICS WITH FIRST-PASSAGE TIME OF STATES AS INDEPENDENT THERMODYNAMIC PARAMETER

V. V. Ryazanov

Institute for Nuclear Research, Kiev, Ukraine, e-mail: vryazan19@gmail.com

**Abstract**

Non-equilibrium states of a thermodynamic statistical system are investigated using the thermodynamic parameter of the system's lifetime, first-passage time, the time before degeneration of the system under influence of fluctuations. Statistical distributions that describe the behavior of energy and lifetime are used. Entropy and obtained thermodynamic relations are compared with the results of Extended Irreversible thermodynamics, where as an additional parameter selected fluxes. Explicit expressions are obtained for average lifetime and conjugate thermodynamic quantity. It is shown that, when is only one stationary non-equilibrium state and exponential distribution for the lifetime, flows can only reduce the average system lifetime. However, there are possibilities to description of increase the average lifetime of the system. A description of the growth of the average lifetime for one stationary non-equilibrium state is possible when choosing distributions for the lifetime, differing from the limiting exponential distribution.

**Introduction**

Limitations appear in classical linear non-equilibrium thermodynamics [1] in the description of such phenomena as the propagation and absorption of ultrasound in liquids, the density profile of shock waves in gases, etc. Attempts to overcome these limitations led to the creation of Extended Irreversible thermodynamics [2-4], in which a variable, additional to the conserved values, the values of the flows are selected. The need to select additional parameters related to the deviation of the system from equilibrium was noted in [5].

Recently stochastic non-equilibrium thermodynamics has been developing [6], where, as in the present work, the finiteness of the observed time interval is taken into account. Informational statistical thermodynamics in [7, 8] was developed based on the method of the non-equilibrium statistical operator (*NSO*) Zubarev [9-10]. The relation between *NSO* and the distributions used in this paper was noted in [11]. In [12, 13, 14, 15] is introduced a distribution that contains the lifetime of the system as an additional thermodynamic parameter. The distributions for the lifetime are described by the equations [16-17], the conjugate kinetic equations for the energy distribution of the system, the main variable of statistical thermodynamics. The value of energy occupies a special place among other physical quantities in the statistical description of equilibrium systems. It can be assumed that the lifetime of the system (first-passage time) plays a similar role in the non-equilibrium case.

Research in finite time spans allows us to draw an analogy with finite-time thermodynamics [18-21], although the latter does not examine precisely the lifetime - the main object of study of this work.

In the present work, as in [12, 13, 14, 15], it is proposed to choose an additional parameter related to the deviation of the system from equilibrium [2-4, 5], in the form of the system lifetime. The lifetime (or first-passage time) is defined as the random moment of the first achievement of a certain boundary (1), for example, of the zero level by a stochastic process that describes the macroscopic parameter $y(t)$, where $y(t)$ is the order parameter of the system, for example, its energy, number of particles, (in some papers, entropy production is considered),

$$\Gamma_x = \inf\{t: y(t)=0\}, \quad y(0)=x>0. \qquad (1)$$

The lifetime in terms of stochastic theory [22] is a functional of the main random process $y(t)$, a subordinate random process.

Stratonovich used the term "lifetime" for the first-passage time in [23]. A number of synonyms are used to characterize this quantity: the first-passage time (for some given level), escape time, busy period (in the queuing theory), etc. The equations for the distribution of the moment of first achievement (lifetime) were obtained in [16]. Kramers in [24] described the escape time from the potential well. Van Kampen [25], Gardiner [26] and other authors [27-34] are discussed the theory of moments of first achievement.

In [12-15] the thermodynamic parameter of the lifetime was introduced; relations for distributions with the lifetime and non-equilibrium entropy were written. But explicit expressions were not obtained for the Laplace transform of the distribution of the system lifetime (first-passage time), the average value of the lifetime and the thermodynamically conjugated parameter. Through this parameter, associated with a change in the entropy of the system during its transition to stationary non-equilibrium state, the average lifetime in stationary non-equilibrium state is expressed. Explicit expressions obtained in the present work, the main motivation of which is precisely the study of the value of the lifetime, which is of considerable interest in various fields of science. We examine the lifetime in a system with flows, under the influences that change the lifetime. The lifetime in stationary non-equilibrium states, the entropy of which, expressed through flows, differs from the equilibrium entropy, also differs from the equilibrium lifetime of an open system. In addition to determining explicit expressions for the lifetime and the conjugated thermodynamic parameter, the task is to find ways to increase the average lifetime.

The lifetime of the system is associated with the time of the stable existence of the system, its response to external and internal influences, with the stability and adaptive abilities of the system. The characteristics of the lifetime $\Gamma$ depend on the energy $u=y(t)$. It is important to give physical interpretation of the mathematical definition (1).The value of lifetime $\Gamma$ is related to the exit time from a set of potential minima. The lifetime is the time before degeneration of the system under influence of fluctuations, associated with the time of stationary existence of the system. The finiteness of the lifetime, for example, the energy of a system is explained by the openness of the system and the stochastic nature of the behavior of the energy. If we set in (1) not the time to reach the zero level, but the time to reach some stationary level, then we can determine the relaxation, correlation, and other characteristic times of the system. In the equilibrium case, in an isolated system, nothing happens with the system or the changes are infinitely slow.The coincidence of lifetime of system to time before the complete termination of their existence depends on force of internal interactions in system, its ability to form structure. In physics, the lifetime usually is understood as time of life of the excited states (nuclei, power levels etc.), i.e. time of existence of random states which are distinct from equilibrium. In our approach lifetime is understood as complete lifetime, including equilibrium and non-equilibrium states, although the focus is on non-equilibrium states. We are considering an open system with fluctuations. The equilibrium lifetime $\Gamma_0$ in such a system is finite. Based on the queuing theory and statistical physics, we can write the expression $\Gamma_0 \sim 1/\lambda P_0 = Q/\lambda$ where $P_0=1/Q$ is the probability of degeneracy, $Q$ is the partition function of the large canonical ensemble, $\lambda=1/\tau$ is the rate of income to the open system in dynamic equilibrium, in a situation of detailed balance [26], $\tau$ is the interval time between arrivals. Lifetime is the maximum time period that includes all other time characteristics of the system. It is possible to formulate mathematically strictly the concept of lifetime of complex systems, understanding under it the time, during which there are elements in the system. The physical value of lifetime of system is introduced in consideration, as an interval of time, during which the system contains a non-zero number of elements, of which the statistical system is made. This value depends on both internal properties of system, and external influences and is generally of random character. The lifetime is connected to characteristic time intervals of system (time of collisions, time of mixing, relaxation time etc.) and with fluxes (also by temporary characteristics). The mathematical party of introduction of lifetime consists in obtaining of the additional information about stochastic process, except for knowledge of its stationary distribution, on stationary properties marked out by the subordinated process. The irreversibility in this approach occurs as a consequence of the assumption about existence and finiteness of lifetime of system, choosing of the random moments of birth and destruction of system. The physical values, for example, heat capacity of system, can essentially depend on lifetime. The introduction of lifetime as thermodynamic parameter is justified by the empiric fact that real systems have finite lifetime, which essentially influences their properties and properties of their environment.

The lifetime of the system may be a very important parameter in technology (the lifetime of electric batteries, for instance), in biology (the lifetime of a biological species, for instance) or in unstable systems. Identifying the parameters affecting the lifetime, lengthening the lifetime, or maximizing the efficiency of the system subject to the restrictions of a finite lifetime seem very interesting topics (some of which may be analogous to some topics arising in finite-time thermodynamics [18-21], in which temporal restrictions on the duration of the cycles or other processes play a central role).

Examples of the use of the lifetime parameter include frequency-tuning systems [17], problems of crossing a potential barrier, decay of an unstable state [26], conformational changes of proteins, ecological systems, epidemics, diffusion-controlled chemical reaction, autocatalytic reaction, number of individuals in a population, dissociation of a diatomic molecule, potential of a neuron, transition through a potential barrier [25]. In [23] the lifetime, the achievement of boundaries by a random process is used to study the emission of random functions, the effects of noise on an electronic relay, and other radio engineering tasks. In [12, 35] the possibilities of applying distributions with a lifetime to the description of superstatistics and such phenomena as the behavior of systems with multiplicative noise, Van der Pol-Duffing systems, turbulence, Van der Pol generators, spatial diffusion of Brownian particles in an external field, the Malthus-Verhulst processare noted.

The definition (1) of the random moment of the first achievement of a certain boundary, as the lifetime and the thermodynamic variable, makes it possible to use not only thermodynamic methods for studying this important quantity, but also stochastic methods and methods of statistical physics. This possibility is important because, as noted in [25], fluctuations cannot be calculated in a system under external influence if the force is specified only macroscopically. It is also necessary to know its stochastic properties. In this work, to study the lifetime, the non-equilibrium entropy of distribution (4) is studied (introduced in [12-15]), and introduced the non-equilibrium partition function, non-equilibrium free energy, and non-equilibrium internal energy.

A rigorous microscopic justification of the proposed statistical approach is fraught with a number of difficulties. Therefore, a macroscopic phenomenological approach is used, as, for example, in [36]. The system is open, and the parameters of the system under the influence of the environment become random variables that are characterized by a stochastic description. In the present work, as in [2–4], an additional thermodynamic parameter is introduced to describe the non-equilibrium state. Therefore, the justification can be carried out, as in [2-4]. In [37] results of works [38, 39] are applied to a substantiation of introduction of the distributions containing time of the first achievement of a level [12-15, 35], lifetime in terminology [23]. The justification can also be carried out using the methods of queuing theory and stochastic storage theory, which consider the distribution of the system's busy period, which can be interpreted as the system's lifetime for small time intervals.

To validate the choice of lifetime as thermodynamic value one can reassert to the method of Zubarev *NSO* [9, 10] which is interpreted in [11] as averaging of quasi-equilibrium (relevant) the statistical operator on distribution of lifetime of the system $p_q(y)$. Then at $p_q(y)=\varepsilon exp\{-\varepsilon y\}$ (in[11-13] a more general choice, merely arbitrary one, of the function $p_q(y)$ is suggested)

$$ln\rho(t)=\int_0^\infty p_q(y)ln\rho_q(t-y,-y)dy=ln\rho_q(t,0)-\int_0^\infty exp\{-\varepsilon y\}(dln\rho_q(t-y,-y)/dy)dy, \qquad (2)$$

where $\rho$ is non-equilibrium statistical operator, $\rho_q$ is relevant statistical operator [9, 10]. The entropy production operator [9, 10] is equal $\sigma(t-u, -u)=dln\rho_q(t-u, -u)/du$. If $\sigma(t-u, -u)\approx\sigma(t)$ or $\sigma(t-u,-u)\approx<\sigma(t)>$ has weak $u$-dependence then equality (3) acquires a form $ln\rho(t)=ln\rho_q(t,0)+<\sigma(t)>\varepsilon^{-1}=ln\rho(t)=ln\rho_q(t,0)+<\sigma(t)><\Gamma>$, as in interpretation [11] $1/\varepsilon=<\Gamma>=<t-t_0>$ is average lifetime. We shall note, that such distribution is received in [40].

Then a relation (2) in non-equilibrium thermodynamics with lifetime as thermodynamic parameter is replaced with expressions (4)-(6), or, in more general form, $ln\rho(t)=ln\rho_q(t,0)+ln[Z(\beta)/Z(\beta,\gamma)]-\gamma\Gamma$, where $\rho=exp\{-\beta E-\gamma\Gamma\}/Z(\beta,\gamma)$; $\rho_q=exp\{-\beta E\}/Z(\beta)$ (in (4) instead of $<\Gamma>(dln\rho_q(t-y_1,-y_1)/dt)$ stand the random value $\Gamma$ and Lagrange multiplier $\gamma$ in: $\gamma\Gamma+lnZ(\beta,\gamma)/Z(\beta)$). That is, averaging on distribution of lifetime in Zubarev *NSO* is replaced with use of a random variable of lifetime, - there lays essential difference of distribution (4)-(6) from (2). In [41], using operator methods, the average lifetime is expressed in terms of a non-equilibrium statistical operator. But the resulting expressions are very cumbersome. In [41] obtained, what is the value of average lifetime decreases or increases in comparison with undisturbed lifetime, depending on a correlation between amount entropy produced in system and environment.

An open system is considered, dynamic quantities under the influence of interaction with the environment become random; only a macroscopic description is possible. When considering systems of finite sizes, it is essential to take into account the finiteness of their lifetime. A normally functioning system is in a stationary non-equilibrium state, characterized by a given deviation from equilibrium and the production of entropy. Each state of the system corresponds to its specific lifetime, associated with the magnitude of the flows and sources in the system and its deviation from equilibrium. The effects that the system is exposed to when interacting with the environment cause deviations from the normal steady state and dissipative effects, change the entropy of the system, and its lifetime.The lifetime is introduced as a phenomenological macroscopic quantity. The lifetime is a statistical value with known equations for the probability density distribution of the lifetime. The introduction of a lifetime as a fundamental thermodynamic random variable seems to be an important and necessary element for the correct description of non-equilibrium phenomena.

This work focuses on the possibilities of increasing the average lifetime (although a more general expression is obtained for the Laplace transform of the distribution of the lifetime). When, under some kind of influence, the basic random process for the energy of the system changes, then the subordinate random process for the random lifetime changes. In the work, these changes are expressed in terms of the parameter $\gamma$ conjugate to the lifetime, which is associated with a change in the entropy of the system.

Section 2 contains distributions describing a system in a stationary non-equilibrium state with a thermodynamic parameter of the lifetime.

In the third section, the entropy and its differentials are determined both for the distribution depending on the energy $y=u$ and the lifetime $\Gamma$, when the specific entropy is equal to $s_\Gamma$ (7), and for the distribution depending only on $y=u$. It is assumed that temporary changes in the entropy of $s_\Gamma$ are caused by flows of entropy entering the system. Then the thermodynamic force $\gamma$ conjugated to the value $\Gamma$ is connected with the flows flowing through the system in a stationary non-equilibrium state. The results are compared with Extended Irreversible thermodynamics [2-4], where the fluxes values are chosen as an additional thermodynamic parameter.

Explicit expressions for the average lifetime $\Gamma$ and the thermodynamic parameter $\gamma$ that is conjugated to $\Gamma$ value are obtained. For the exponential distribution for the lifetime, flows can only reduce the average lifetime.

In the fourth section, we consider a system of a random number of non-equilibrium stationary states. For the non-exponential distribution of the lifetime, a result significantly different from one stationary non-equilibrium state of exponential distribution of the lifetime is obtained: conditions for changes in entropy under which an increase in the average lifetime occurs.

In conclusion, the results are discussed.

## 2. Distributions for energy and lifetime

The lifetimes are affected by attractors, metastable states, phase transitions, and other physical features of the system that depend on dynamic variables ($z=q_1,...,q_N, p_1,...,p_N$), $q$ and $p$ are the coordinates and momenta of $N$ particles of the system. The lifetime depends on the energy $u$ and the number of particles $N$, depending on $z$. Moreover, the lifetime is a macroscopic and slowly changing quantity.

The distribution for the lifetime $\Gamma$ (1) depends on the macroscopic values of $y(t)$. Suppose that the process $y(t)=u$ is the energy (or another system order parameter) of the system. The relationship between the distribution density $p(u,\Gamma)=p_{u\Gamma}(x,y)$ and the microscopic (coarse-grained) density $\rho(z;u,\Gamma)$ is written as (the standard procedure, for example, [42])

$$p(u, \Gamma)=\int\delta(u-u(z))\delta(\Gamma-\Gamma(z))\rho(z; u, \Gamma)dz=\rho(u,\Gamma)\omega(u,\Gamma). \quad (3)$$

The factor $\omega(u)$ is replaced by $\omega(u,\Gamma)$, the volume of the hypersurface in the phase space containing fixed values of $u$ and $\Gamma$. If $\mu(u,\Gamma)$ is the number of states in the phase space with parameters less than $u$ and $\Gamma$, then $\omega(u,\Gamma)=d^2\mu(u,\Gamma)/dud\Gamma$. Moreover, $\int\omega(u,\Gamma)d\Gamma=\omega(u)$. The number of phase points with parameters lying in the interval between $u, u+du; \Gamma, \Gamma+d\Gamma$, is equal to $\omega(u,\Gamma)dud\Gamma$.

We now use the principle of equal probabilities for an extended phase space with cells $(u,\Gamma)$ (instead of the principle of equal probability for phase space with cells $(u)$; in this case, realizations of the space of elementary events change). Using the principle of maximum entropy [43], we write the expression for the microscopic probability density in the extended phase space

$$\rho(z;u,\Gamma)=exp\{-\beta u(z)-\gamma\Gamma(z)\}/Z(\beta,\gamma), \quad (4)$$

where

$$Z(\beta,\gamma) = \int e^{-\beta u-\gamma\Gamma}dz = \iint du\ d\Gamma\ \omega(u,\Gamma)e^{-\beta u-\gamma\Gamma} \quad (5)$$

is the partition function, $\beta$ and $\gamma$ are the Lagrange multipliers satisfying the following expressions for the averages:

$$<u>=-\partial lnZ/\partial\beta|_\gamma; \qquad <\Gamma>=-\partial lnZ/\partial\gamma|_\beta. \quad (6)$$

In expressions (4)-(6), the values of energy $u$ and lifetime $\Gamma$ are chosen as thermodynamic parameters. The thermodynamically conjugated lifetime value $\gamma$ as can be seen from (2), is associated with the production and flows of entropy, which characterize the peculiarities of the non-equilibrium processes in an open statistical system. At $\gamma=0$ and $\beta=\beta_0=T^{-1}_{eq}$, where $T_{eq}$ is the equilibrium temperature, the non-equilibrium distribution (4) yield the equilibrium Gibbs distribution. The distribution (4) over a lifetime is a generalization of Gibbs distribution to a non-equilibrium situation. The canonical Gibbs distribution is obtained from the microcanonical ensemble in the zero approximation by the interaction of the system with the environment.

Using the lifetime $\Gamma$, an effective account is taken of this interaction (similarly to the methods of McLennan [44] and Zubarev *NSO* [9, 10]). Physical phenomena such as metastability, phase transitions, attracting attractors and repellers, stationary non-equilibrium states violate the equal probability hypothesis of phase space with cells (*u*) points. The value of $\gamma$ can be considered as measure of deviation from equilibrium. Many works are close to non-equilibrium thermodynamics obtained from the distribution with the lifetime as a thermodynamic parameter [12, 13]. For example:

- Extended Irreversible thermodynamics [2-4], in which flows, quantities inverse to the lifetime are selected as an additional thermodynamic parameter.

- Informational statistical thermodynamics [7, 8], based on the Zubarev *NSO* [9-10], in which an infinite number of thermodynamic parameters, flows of all orders are selected. The truncation operation, the limitation by a finite number of parameters, leads to the fact that the lifetime is also determined with some accuracy. It was noted in [7, 8] that the *NSO* contains a hypothesis of a non-mechanical nature. This is a hypothesis about the possibility of using and finiteness of the system lifetime [11].

## 3. Generalized thermodynamics of systems with parameter of lifetime

In the general case, the value of $\Gamma$ can be selected as a subprocess in a form different from that described above. Then the equilibrium situation will mean some curve in the ($\beta,\gamma$) plane, which does not necessarily coincide with the line $\gamma=0$. Features of the non-analytic behavior of the partition function (or its derivatives) on the ($\beta,\gamma$) plane indicate non-equilibrium phase transitions.

Assuming the measurability of the lifetime, we introduce the local specific entropy $s_\Gamma$ corresponding to distribution (4) (*u* is specific internal energy) by the relation [12-14]

$$s_\Gamma = -\langle \ln\rho(z;u,\Gamma)\rangle = \beta\langle u\rangle + \gamma\langle\Gamma\rangle + \ln Z(\beta,\gamma); \quad ds_\Gamma = \beta d\langle u\rangle + \gamma d\langle\Gamma\rangle. \tag{7}$$

For spatially inhomogeneous systems, the quantities $\beta$ and $\gamma$ in the general case depend on the spatial coordinate. Distributions (4) can be considered for a small volume element in which the values of $\beta$ and $\gamma$ are replaced by the average values constant over this volume element over this volume element. In non-equilibrium thermodynamics, the densities of extensive thermodynamic quantities (entropy, internal energy, mass fraction of a component) are considered. We follow this approach, including here and the lifetime.

The Fourier law $q=-\lambda\nabla T$ (where $q$ is the heat flux, $\lambda$ is the heat conductivity coefficient, $T$ is the absolute temperature) of classical non-equilibrium thermodynamics [1] is replaced in Extended Irreversible thermodynamics [2-4] by the Maxwell-Cattaneo equation of the form

$$q = -\lambda\nabla T - \tau_q \partial q/\partial t, \tag{8}$$

where $\tau_q$ is the correlation time of flows. In this case, the Fourier law is generalized, taking the form:

$$q(t) = -\int_0^t \varphi(t-t`)\nabla T(t`)dt`. \tag{9}$$

The solution of equation (8) is $q(t)=-\int_0^t [(\lambda/\tau)exp\{-(t-t`)/\tau]\nabla T(t`)dt`$, then the memory function is $\varphi(t-t`)=(\lambda/\tau)exp\{-(t-t`)/\tau$. For $\tau\to 0$, $\varphi(t-t`)=\lambda\delta(t-t`)$. Substituting this value of the memory function in (9), we obtain the Fourier equation.

In distribution (4)-(5), containing the lifetime as a thermodynamic parameter, the joint probability (3)-(4) for the quantities $u$ and $\Gamma$ is

$$P(u,\Gamma) = \frac{e^{-\beta u - \gamma \Gamma}\omega(u,\Gamma)}{Z(\beta,\gamma)}. \tag{10}$$

Integrating (10) over $\Gamma$, we obtain a distribution for $u$ of the form

$$P(u) = \int P(u,\Gamma)d\Gamma = \frac{e^{-\beta u}}{Z(\beta,\gamma)}\int_0^\infty e^{-\gamma\Gamma}\omega(u,\Gamma)d\Gamma. \tag{11}$$

The factor $\omega(u,\Gamma)$ is the joint probability for $u$ and $\Gamma$, considered as the stationary probability of this process. We rewrite the value $\omega(u,\Gamma)$ in the form

$$\omega(u,\Gamma) = \omega(u)\omega_1(u,\Gamma) = \omega(u)\sum_{k=1}^n R_k f_{1k}(\Gamma,u). \tag{12}$$

In equality (12) it is assumed that there are $n$ classes of states in the system; $R_k$ is the probability that the system is in the $k$-th class of states, $f_{1k}(\Gamma,u)$ is the density of the distribution of the lifetime $\Gamma$ in this class of (ergodic) states (in the general case, $f_{1k}(\Gamma,u)$ depends on $u$). As a physical example of such a situation (characteristic of metals, glasses, etc.), one can cite the potential of many complex physical systems. Below we restrict ourselves to the case $n=1$.

Such a situation was considered in [45, 46]. The minimum points of the potential correspond to metastable phases, disordered structures, etc. The phase space of these systems is divided into isolated regions, each of which corresponds to a metastable thermodynamic state, and the number of these regions increases exponentially with increasing total number of (quasi)particles [47].

The finiteness of the lifetime of a system implies the finiteness of its size. The thermodynamic limit is not calculated, although the dimensions of the system are assumed so large that it is possible to implement a probabilistic ensemble as the state of the system, and to obtain average values, you can choose not infinite, but very large values of the volume and number of particles with a small error. In the theory of random processes, when the definition (1) is valid, there is the concept of the limit of a sequence of random variables. There are at least four ways to determine this limit. Queuing theory addresses the limits of intervals between receipts and other events.

You can consider finite systems, or you can make the limit transition, as in the theory of random processes. Moreover, the volume of the system can be finite, an infinite number of observations requires infinite time. Infinite time also appears in the definitions of ergodicity and mean equilibrium value. One of the arguments for introducing the thermodynamic limit is the elimination of boundary effects (for example, [48]). However, for the finiteness of the lifetime, the existence of a boundary is necessary. In [49] large values of the number of particles $N$ are considered, but there is no transition to the limit.

To obtain an explicit form of the distribution in (12), we use the general results of the mathematical theory of phase enlargement of complex systems [50], from which the following exponential form of the density of the distribution of the lifetime for an enlarged random process follows (see also [23]):

$$p(\Gamma < y) = \Gamma_0^{-1} \exp\{-y/\Gamma_0\} \tag{13}$$

for one class of stable states, and Erlang distribution density

$$p(\Gamma < y) = \sum_{k=1}^n R_k \Gamma_{0k}^{-1} \exp\{-y/\Gamma_{0k}\}; \qquad \sum_{k=1}^n R_k = 1 \tag{14}$$

in the case of several ($n$) classes of ergodic stable states, $\langle \Gamma_{\gamma k} \rangle$ is the average lifetime in the $k$-th class of metastable states with a perturbation $\gamma = \gamma_k$, $\Gamma_{0k} = \langle \Gamma_{0k} \rangle$ is the average lifetime in the $k$-th class of metastable states without a perturbation $\gamma = \gamma_k = 0$, in some basic stationary (or equilibrium) state. Do not confuse the notation $\Gamma_0$ with the notation from (1) for $x=0$.

The exponential form of distribution (13) can be justified using, for example, the principle of maximum entropy, general approaches of statistical physics, and also using approaches of information geometry [51].

The values $\langle \Gamma_{0k} \rangle$ and $\langle \Gamma_{\gamma k} \rangle$ are the averaging of the residence time and degeneracy probabilities over stationary ergodic distributions (in our case, the Gibbs distribution). The physical reason for the implementation of the distribution in the form of (13)-(14) is the presence of weak ergodicity in the system, which in the limit leads to stationary distributions. The mixing of system states at large times leads to distributions (13)-(14). The structural factor $\omega(u,\Gamma)$ has the value of the joint probability density of the values of $u$, $\Gamma$. For distributions (13), (14), the functions $\omega(u,\Gamma)$ from (12) have the form:

$$\omega(u,\Gamma) = \omega(u) \sum_{k=1}^{n} R_k \Gamma_{0k}^{-1} \exp\{-\Gamma/\Gamma_{0k}\}; \quad \omega(u,\Gamma) = \omega(u)\Gamma_0^{-1}\exp\{-\Gamma/\Gamma_0\}; \quad n=1. \tag{15}$$

The form of the function $p_\Gamma(y)$ (13)-(14) reflects not only the internal properties of the system, but also the influence of the environment on the open system, and the features of its interaction with the environment. A physical interpretation of the exponential distribution for the function $p_\Gamma(y)$ is given in [9, 10]: the system evolves as an isolated system controlled by the Liouville operator. In addition, the system undergoes random transitions, and the phase point representing the system switches from one trajectory to another with an exponential probability under the influence of a "thermostat". Exponential distribution describes completely random systems. The influence of the environment on the system can also be organized in nature, for example, this applies to systems in a non-equilibrium state with input and output unsteady flows. The nature of the interaction with the environment may also change; therefore, various forms of the function $p_\Gamma(y)$ can be used [13].

Note that a value similar to $\gamma$ is determined in [52]–[54] for a fractal repeller object. It is equal to zero for a closed system, and for an open system it is equal to $\Sigma\lambda_i - \lambda_{KS}$, where $\lambda_i$ are Lyapunov exponents, $\lambda_{KS}$ is the Kolmogorov-Sinai entropy. The thermodynamic interpretation of $\gamma$ is given below.

We consider the case when the average lifetime $\Gamma_0$ in (13) depends on a random variable $u$. Then, integrating over $\Gamma$, as in (11), we obtain for $n=1$ from (5), (13), (15) that

$$Z(\beta,\gamma) = \int e^{-\beta u}\omega(u)\frac{du}{1+\gamma\Gamma_0(u)}. \tag{16}$$

In Extended Irreversible thermodynamics [2-4], the differential of the entropy density for the case of thermal conductivity is

$$ds = \theta^{-1}du_\beta - \frac{\tau}{\rho\lambda\theta^2}qdq, \quad S = \int_{V(t)}\rho s dV, \tag{17}$$

where $S$ is entropy, $s$ is local specific entropy, $V(t)$ is volume, $u=u_\beta$ is specific internal energy, $q$ is the heat flux, $\theta^{-1}$ is the non-equilibrium temperature, $\rho$ is the mass density, $\lambda$ is the thermal conductivity, $\tau=\tau_q$ is the correlation time of the fluxes from equation (8),

$$s_{eq} = s_{\Gamma|\gamma=0} = \beta u_\beta + \ln Z_\beta, \quad Z_\beta = \int e^{-\beta u}\omega(u)du, \quad u_\beta = -\partial\ln Z_\beta/\partial\beta, \tag{18}$$

where $Z_\beta$ is equilibrium partition function, $u_\beta$ is equilibrium energy, $s_{eq}$ is equilibrium entropy. The non-equilibrium temperature is

$$\theta^{-1} = T^{-1} - \frac{1}{2}\frac{\partial}{\partial u_\beta}(\frac{\tau}{\rho\lambda\theta^2})q^2, \tag{19}$$

specific entropy $s$ and entropy production $\sigma^s$ are equal

$$s(u,q) = s_{eq}(u) - \frac{1}{2}\frac{\tau}{\rho\lambda\theta^2}q^2, \sigma^s = \frac{1}{\lambda\theta^2}qq. \tag{20}$$

Comparing the expressions for entropy (20) and the differential of entropy (17) from Extended Irreversible thermodynamics with variable values of energy and flows with the same expressions, which include the lifetime obtained from expressions (7), (6), we write the relation

$$ds = \theta^{-1} du_\beta - a_\beta q dq = \beta d\bar{u} + \gamma d\bar{\Gamma} = \beta \frac{\partial \bar{u}}{\partial \beta}\bigg|_{\bar{\Gamma}} d\beta + (\gamma + \beta \frac{\partial \bar{u}}{\partial \bar{\Gamma}}\bigg|_\beta) d\bar{\Gamma}, \quad a_\beta = \frac{\tau}{\rho \lambda \theta^2}, \quad s = s_\Gamma, \quad (21)$$

where the variables $\beta$ and $\Gamma$ are selected on the right-hand side. We rewrite (21) in the form

$$\gamma + \beta \frac{\partial \bar{u}}{\partial \bar{\Gamma}}\bigg|_\beta + \beta \frac{\partial \bar{u}}{\partial \beta}\bigg|_\Gamma \frac{\partial \beta}{\partial \bar{\Gamma}}\bigg| = \theta^{-1} \frac{\partial u_\beta}{\partial \bar{\Gamma}}\bigg| - a_\beta q \frac{\partial q}{\partial \bar{\Gamma}}\bigg| \quad . \quad (22)$$

Given a constant value of $\gamma$ in relation (22), we obtain

$$\gamma = \frac{1}{\frac{\partial \bar{\Gamma}}{\partial \beta}\bigg|_\gamma}(\theta^{-1}\frac{\partial u_\beta}{\partial \beta}\bigg|_\gamma - \beta \frac{\partial \bar{u}}{\partial \beta}\bigg|_\gamma) \quad . \quad (23)$$

Expression (23) is also written from (22) and when a constant value of $q$ is set. Setting constant values of $\beta$ or $u_\beta$ in (22), we obtain

$$\gamma = -\frac{1}{\frac{\partial \bar{\Gamma}}{\partial \gamma}\bigg|_\beta}(\beta \frac{\partial \bar{u}}{\partial \gamma}\bigg|_\beta + a_\beta q \frac{\partial q}{\partial \gamma}\bigg|_\beta) \quad . \quad (24)$$

From (24) and (23) we obtain

$$\frac{\partial q}{\partial \gamma}\bigg|_\beta = \frac{1}{a_\beta q}(-\gamma \frac{\partial \bar{\Gamma}}{\partial \gamma}\bigg|_\beta - \beta \frac{\partial \bar{u}}{\partial \gamma}\bigg|_\beta) \quad , \quad \beta \frac{\partial \bar{u}}{\partial \beta}\bigg|_\gamma = \theta^{-1} \frac{\partial u_\beta}{\partial \beta}\bigg|_\gamma - \gamma \frac{\partial \bar{\Gamma}}{\partial \beta}\bigg|_\gamma \quad . \quad (25)$$

For the time derivative of the average lifetime, expressions are written

$$\gamma \frac{d\bar{\Gamma}}{dt} = \theta^{-1} \frac{du_\beta}{dt} - a_\beta q \frac{dq}{dt} - \beta \frac{d\bar{u}}{dt} \quad ;$$

$$(\gamma + \beta \frac{\partial \bar{u}}{\partial \bar{\Gamma}}\bigg|_\beta) \frac{d\bar{\Gamma}}{dt} = \theta^{-1} \frac{du_\beta}{dt} - a_\beta q \frac{dq}{dt} - \beta \frac{\partial \bar{u}}{\partial \beta}\bigg|_{\bar{\Gamma}} \frac{d\beta}{dt} \quad ; \quad (26)$$

$$(\gamma + \beta \frac{\partial \bar{u}}{\partial \bar{\Gamma}}\bigg|_\gamma) \frac{d\bar{\Gamma}}{dt} = \theta^{-1} \frac{du_\beta}{dt} - a_\beta q \frac{dq}{dt} - \beta \frac{\partial \bar{u}}{\partial \gamma}\bigg|_\Gamma \frac{d\gamma}{dt} \quad ;$$

$$(\gamma + \beta \frac{\partial \bar{u}}{\partial \bar{\Gamma}}\bigg|_q) \frac{d\bar{\Gamma}}{dt} = \theta^{-1} \frac{du_\beta}{dt} - a_\beta q \frac{dq}{dt} - (\beta \frac{\partial \bar{u}}{\partial q}\bigg|_{\bar{\Gamma}} + \gamma \frac{\partial \bar{\Gamma}}{\partial q}\bigg|_{\bar{\Gamma}}) \frac{dq}{dt} \quad ; \quad (27)$$

$$(\gamma + \beta \frac{\partial \bar{u}}{\partial \bar{\Gamma}}\bigg|_{u_\beta}) \frac{d\bar{\Gamma}}{dt} = \theta^{-1} \frac{du_\beta}{dt} - a_\beta q \frac{dq}{dt} - (\beta \frac{\partial \bar{u}}{\partial u_\beta}\bigg|_{\bar{\Gamma}} + \gamma \frac{\partial \bar{\Gamma}}{\partial u_\beta}\bigg|_{\bar{\Gamma}}) \frac{du_\beta}{dt} \quad .$$

Using expressions (6), (16), we write the value $\gamma\langle\Gamma\rangle$ in the relation for specific entropy (7) in the form:

$$s_\Gamma = s = -\langle \ln\rho(z;u,\Gamma)\rangle = \beta\langle u\rangle + \gamma\langle\Gamma\rangle + \ln Z(\beta,\gamma) =$$

$$\int e^{-\beta u}\omega(u)(\frac{x}{(1+x)^2})du/Z + \beta\langle u\rangle + \ln Z(\beta,\gamma), \quad x = \gamma\Gamma_0(u), \quad (28)$$

and we equate the quantity $\Delta s = -\Delta = s_\Gamma - s_{eq}$ and (18) to the corresponding expression of the Extended Irreversible thermodynamics (20) [2-4], where this value is equal $-a_\beta q^2/2$, $a_\beta = \tau/\rho\lambda T^2$. The value $Z$ is equal to (16). And for the average lifetime in (28), the expression obtained from (16) and (6) is used

$$<\Gamma> = -\frac{\partial \ln Z}{\partial \gamma}\bigg|_\beta = \frac{1}{Z}\int e^{-\beta u}\omega(u)\frac{\Gamma_0 du}{(1+x)^2} \quad . \tag{29}$$

We rewrite expressions (28), (29) in the form

$$Z\gamma\overline{\Gamma} = \int e^{-\beta u}\omega(u)(\frac{x}{(1+x)^2})du = Z[\beta(u_\beta - \overline{u}) - a_\beta q^2/2 - \ln Z(\beta,\gamma)/Z_\beta]. \tag{30}$$

We assume that the fluxes $q$ weakly depend on $\beta$ and neglect this dependence (also in [2]). The internal energy $u_\beta$ does not depend on $\gamma$ and $q$. Thus, in this approximation, the fluxes $q$ depend only on $\gamma$, and $u_\beta$ depends only on $\beta$.

The initial relation (28) has the form

$$s_\Gamma = s = s_{eq} - a_\beta q^2/2 = \gamma\overline{\Gamma} + \beta\overline{u} + \ln Z = \beta u_\beta + \ln Z_\beta - a_\beta q^2/2. \tag{31}$$

From relations (16) and (6) we obtain (29) and

$$<u> = -\frac{\partial \ln Z}{\partial \beta}\bigg|_\gamma = \frac{1}{Z}\int e^{-\beta u}\omega(u)[\frac{u}{1+\gamma\Gamma_0(u)} + \frac{d(\gamma\Gamma_0)/d\beta}{(1+\gamma\Gamma_0(u))^2}]du \quad . \tag{32}$$

We substitute the expressions (29) and (32) in (31), multiply by $Z$ and equate the integrands. We get the expression

$$\beta\frac{dx}{d\beta} + x[1-\beta(u_\beta-u)+\ln Z/Z_\beta + a_\beta q^2/2] - \beta(u_\beta-u) + \ln Z/Z_\beta + a_\beta q^2/2 = 0. \tag{33}$$

Differentiating (31) with respect to $\beta$, we obtain

$$\gamma\frac{\partial\overline{\Gamma}}{\partial\beta} + \beta\frac{\partial(\overline{u}-u_\beta)}{\partial\beta} + \frac{1}{2}q^2\frac{\partial a_\beta}{\partial\beta} = 0. \tag{34}$$

If now in (31) we take into account dependence (16) only in $lnZ$, differentiate with respect to $\beta$, multiply by $Z$ and equate the integrands, then for $dx/d\beta$, taking into account (34), we obtain

$$\beta\frac{dx}{d\beta} - x\beta(\overline{u}-u) - \beta(\overline{u}-u) = 0. \tag{35}$$

Comparing this expression (35) with (33), we obtain

$$x = \frac{a_3}{1-a_3}, \quad a_3 = \beta(u_\beta - \overline{u}) - \ln Z/Z_\beta - a_\beta q^2/2. \tag{36}$$

It can be seen that the right-hand side of (36) is independent of random variables. Therefore, averaging applies only to the left side of expression (36). Averaged over the equilibrium distribution and below we mean by the value $x = \gamma\overline{\Gamma}_0(u)$ this value averaged over the equilibrium distribution. We do not use the non-equilibrium distribution (4), since $\Gamma_0(u)$ it does not depend on $\gamma$. We assume that the average value is independent of $\gamma$. Substituting this expression (36) into (16), we obtain, taking into account (31), that

$$Z = Z_\beta[1-\beta(u_\beta-\overline{u})+\ln Z/Z_\beta + a_\beta q^2/2] = Z_\beta[1-\gamma\overline{\Gamma}], \tag{37}$$

i.e., relation (41).

Differentiating (37) with respect to $\gamma$, we obtain, taking into account (6)

$$\gamma\langle\Gamma\rangle^2 + \gamma\frac{\partial\langle\Gamma\rangle}{\partial\gamma} = 0. \tag{38}$$

From this differential equation with the initial condition $\langle\Gamma\rangle_{\gamma=0}=\Gamma_0$ we obtain expression (39) for the average lifetime

$$\langle\Gamma\rangle = \frac{\Gamma_0}{1+x}. \tag{39}$$

From this and from (37) we obtain that $Z = Z_\beta Z_\gamma$, $Z_\gamma = 1/(1+x)$. The same result is obtained by substituting (13) into the relation

$$Z_\gamma = \int_0^\infty e^{-\gamma \Gamma} p(\Gamma) d\Gamma, \qquad (40)$$

where $p(\Gamma)$ is the probability density of the distribution of the lifetime of the form (13). It also follows that the value of the unperturbed lifetime $\Gamma_0(u)$ does not depend on the random variable $u$, but on the average value of $u_\beta$ and the parameter $\beta$. There is a separation of the partition function into equilibrium and non-equilibrium factors. In this case, free energy $-\beta^{-1} \ln Z$ is divided into two terms, $-\beta^{-1} \ln Z = -\beta^{-1} \ln Z_\beta - \beta^{-1} \ln Z_\gamma$ where $-\beta^{-1} \ln Z_\beta$ is the equilibrium part of free energy, and $-\beta^{-1} \ln Z_\gamma$ is the non-equilibrium part of free energy, and $-\beta^{-1} \ln Z_{\gamma|\gamma=0} = 0$; $\bar{u} = u_\beta + u_\gamma$, $u_{1\gamma} = -\dfrac{\partial \ln Z_{1\gamma}}{\partial \beta}$ is non-equilibrium energy. We also note that the quantity $Z_\gamma = \int_0^\infty e^{-\gamma \Gamma} p(\Gamma) d\Gamma$ (40) has probabilistic meaning: $Z_\gamma = \int_0^\infty e^{-\gamma \Gamma} p(\Gamma) d\Gamma = P\{\Gamma \le \tau\}$; $P\{\tau > t\} = \exp\{-\gamma t\}$. At $\gamma \sim \sigma_s$, $\sigma_s$ is entropy production, $P\{\tau > t\} \sim 1$, when $\sigma_s \to 0$, and $P\{\tau > t\} \sim 0$, when $\sigma_s \to \infty$; $P\{\Gamma \le \tau\} \sim 1$, $\tau \to 0$.

From (31) and (39) we obtain

$$\frac{Z}{Z_\beta} \approx (1+\gamma \Gamma_{0\beta})^{-1} = e^{\ln Z/Z_\beta} = 1 + \beta(\bar{u} - u_\beta) + \ln Z/Z_\beta + a_\beta q^2/2 = 1 - \gamma \bar{\Gamma}, \quad \bar{\Gamma} = \frac{\Gamma_{0\beta}}{1+\gamma \Gamma_{0\beta}}, \quad \Gamma_{0\beta} = \Gamma_0. \qquad (41)$$

Differentiating (41) with respect to $\beta$, we obtain

$$\frac{Z}{Z_\beta}(u_\beta - \bar{u}) = \frac{1}{2} q^2 \frac{da_\beta}{d\beta} + \beta \frac{\partial(\bar{u} - u_\beta)}{\partial \beta}. \qquad (42)$$

Differentiating (37) with respect to $\gamma$, we obtain, taking into account the equality $\dfrac{\partial \langle \Gamma \rangle}{\partial \beta} = \dfrac{\partial \langle u \rangle}{\partial \gamma}$, that

$$\gamma \langle \Gamma \rangle^2 = \beta \frac{\partial \langle \Gamma \rangle}{\partial \beta} + a_\beta q \frac{dq}{d\gamma}. \qquad (43)$$

From (43) and (41)

$$\frac{\partial \langle \Gamma \rangle}{\partial \beta} = \frac{d\Gamma_0}{d\beta} \frac{1}{(1+x)^2}, \quad x = \gamma \Gamma_0, \quad \beta \frac{d\Gamma_0}{d\beta} = \gamma \Gamma_0^2 - a_\beta q \frac{dq}{d\gamma}(1+x)^2. \qquad (44)$$

From (31), (41), differentiating (41) with respect to $\beta$, we obtain

$$\bar{u} - u_\beta = u_\gamma = \gamma \frac{d\Gamma_0}{d\beta} \frac{1}{1+x}, \qquad (45)$$

and taking into account (43) we obtain a differential equation for $a_\beta q^2/2$ of the form

$$\gamma(1+x) d(a_\beta q^2/2)/d\gamma - a_\beta q^2/2 = x - \ln(1+x)$$

with the decision

$$a_\beta q^2/2 = \ln(1+x). \qquad (46)$$

Then

$$\gamma = \frac{1}{\Gamma_0}(e^{a_\beta q^2/2} - 1), \quad 1 + x = e^{a_\beta q^2/2}, \quad \langle \Gamma \rangle = \frac{\Gamma_0}{1+x} = \Gamma_0 e^{-a_\beta q^2/2}. \qquad (47)$$

Heat fluxes reduce lifetime. From (47) we write the relations for the derivatives

$$a_\beta q\gamma \frac{dq}{d\gamma} = \frac{x}{1+x}, \quad \frac{1}{2}q^2\beta \frac{da_\beta}{d\beta} = -\frac{x}{1+x} = -\gamma\langle\Gamma\rangle. \tag{48}$$

For non-equilibrium temperature (19), taking into account (48), we obtain

$$\theta^{-1} = \beta - \frac{1}{2}q^2 \frac{da_\beta}{d\beta}\frac{d\beta}{du_\beta} = \beta + \frac{x}{1+x}\frac{1}{\beta du_\beta/d\beta}. \tag{49}$$

The derivatives of the lifetime from (43), (44), taking into account (48), are equal

$$\beta\frac{d\overline{\Gamma}}{d\beta} = -\frac{\overline{\Gamma}}{1+x}, \quad \beta\frac{d\Gamma_0}{d\beta} = -\Gamma_0, \quad \frac{d\overline{\Gamma}}{d\gamma} = -\langle\Gamma\rangle^2. \tag{50}$$

Using expressions (47), (48), (50) in (49), we find

$$\theta^{-1}\frac{du_\beta}{d\beta} = \beta\frac{du_\beta}{d\beta} + \frac{\gamma\overline{\Gamma}}{\beta}, \quad \theta^{-1} = \beta - \frac{(\overline{u}-u_\beta)}{du_\beta/d\beta} = \beta + \frac{(1-e^{-a_\beta q^2/2})}{\beta du_\beta/d\beta}. \tag{51}$$

For the energy deviation from the equilibrium value, we obtain

$$\overline{u} - u_\beta = u_\gamma = -\frac{1}{\beta}\frac{x}{1+x} = -\frac{\gamma\overline{\Gamma}}{\beta} = \frac{1}{\beta}(e^{-a_\beta q^2/2} - 1). \tag{52}$$

Integrating expression (50), we obtain

$$\Gamma_0 = \frac{C_\Gamma}{\beta} = C_\Gamma T, \tag{53}$$

where $C_\Gamma$ is independent of $\beta$, $\gamma$, $u_\beta$, $q$. From (50), (52) we find

$$\frac{d\overline{u}}{d\beta} = \frac{du_\beta}{d\beta} + \frac{\gamma\overline{\Gamma}}{\beta^2}(1+\frac{\overline{\Gamma}}{\Gamma_0}), \quad \frac{d\overline{u}}{d\gamma} = \frac{\overline{\Gamma}}{\beta}(\gamma\overline{\Gamma}-1). \tag{54}$$

The flows and their derivatives are equal

$$q = [\frac{2}{a_\beta}\ln(1+x)]^{1/2}, \quad \frac{dq}{d\gamma} = \frac{1}{[2a_\beta\ln(1+x)]^{1/2}}\frac{\Gamma_0}{1+x} = \frac{\Gamma_0}{a_\beta q}e^{-a_\beta q^2/2}. \tag{55}$$

Expressions (26), (27) are written in more detail.

In the comments after expression (2), it was noted that the parameter $\gamma$, which is the lifetime conjugate, is associated with the production of entropy. From (47) we obtain

$$\gamma = \frac{1}{\Gamma_0}(e^{a_\beta q^2/2} - 1) \approx \frac{-\Delta s}{\Gamma_0}, \quad \Delta s = s - s_{eq} = -\Delta = -a_\beta q^2/2. \tag{56}$$

Thus, the production of entropy coincides with the parameter $\gamma$ only in the first approximation of the expansion of the exponent in a series, and not as in expression (2). But in (47) and (56), it is precisely the nonlinearity of the quantity $\gamma$ conjugate with the lifetime that is important.

A number of expressions for the non-equilibrium temperature (49), (51), average lifetime (39), (41), (47), and also other expressions for the average lifetime are written down, for example, using expressions (45), (47), (48), (50), (52), we obtain

$$\overline{\Gamma} = \Gamma_0 \frac{\beta(\overline{u}-u_\beta)(1+q^2\frac{1}{2}\beta\frac{\partial a_\beta}{\partial\beta})}{q^2\frac{1}{2}\beta\frac{\partial a_\beta}{\partial\beta}} = -q^2\frac{1}{2}\beta\frac{\partial a_\beta}{\partial\beta}\frac{1}{\gamma} = a_\beta q\frac{\partial q}{\partial\gamma} = -\Gamma_0\frac{\beta(\overline{u}-u_\beta)}{(e^{-\Delta s}-1)}. \tag{57}$$

The obtained relations are generalized to the cases of the presence of other influences on the system, except for the heat flux. For example, if there is still viscous pressure $P^{0v}$, then expression (46) takes the form $a_\beta q^2/2 + (\tau_2/4\eta T)P^{0v} : P^{0v} = \ln(1+x_1)$, where $\eta$ is the viscosity. We will consider arbitrary changes in entropy and denote it $\Delta_1 = -\Delta s = s_{eq} - s$.

From (31) and (47) we obtain

$$\langle\Gamma\rangle = \frac{\Gamma_0}{1+x} = \Gamma_0 e^{-a_\beta q^2/2} = \Gamma_0 e^{\Delta s}, \quad \Delta s = -\frac{1}{2}a_\beta q^2 \leq 0. \tag{58}$$

Any flows for one stationary non-equilibrium state and exponential distribution for the lifetime reduce the lifetime. This applies to substance flows, viscous pressure, and other flows. The same results can be obtained from a comparison with the results of the *NSO* [9, 10] and informational statistical thermodynamics [7, 8].

This situation can be visualized so that the energy under the influence will more intensively pass through the "potential barrier" to the state of equilibrium from the "potential well".

In this paper, as in [2], the case is considered, when $\Delta s = s - s_{eq} = -\Delta_1 \leq 0$. However, for example, in [55] it was shown that in a stationary non-equilibrium state negative entropy fluxes enter the system, and in the general case a situation is possible when $\Delta s > 0$. Then the value $\gamma$ in (56) will be negative, and the value $\langle \Gamma \rangle$ in (58) will be greater than $\Gamma_0$.

## 4. **Possibilities for describing of several stationary non-equilibrium states. Possibilities of increasing lifetime**

The exponential distributions for the lifetime were considered above. We now consider the complex Poisson distribution for the lifetime corresponding to a random number of stationary non-equilibrium states and potential wells. We proceed from the general expression (12) for *n*=1, and for the distribution of the lifetime we choose a complex Poisson distribution with the Laplace transform of the form

$$\ln \varphi(\gamma) = -a(1 - p(\gamma)). \tag{59}$$

The generating function of Poisson distribution is
$$\varphi(z) = e^{-a(1-z)}.$$

If the argument $z$ in this expression is replaced by the Laplace transform of the exponential distribution $p(\gamma) = \dfrac{1}{1 + \gamma \Gamma_{01}}$, $x = \gamma \Gamma_{01}$, then we get the Laplace transform of the distribution describing the sum $\eta = \sum_{k=1}^{\nu} \zeta_k$ independent of each other and independent of $\nu$, identically distributed with an exponential distribution of random variables with Laplace transform $p_\eta(\gamma) = \varphi(p(\gamma))$, where $\nu$ is non-negative integer random variable with generating function $\varphi(z) = e^{-a(1-z)}$. So we will describe the sum of a random number of potential wells with an exponential distribution of the system lifetime in each potential well; the value of *a* represents the average number of potential wells in this sum. Such relations are valid not only for the complex Poisson distribution. If the statistical sum is written as $Z(\beta, \gamma) = Z_\beta Z_\gamma$, then part of the statistical sum $Z_\gamma = \int e^{-\gamma \Gamma} p(\Gamma) \, d\Gamma$ (40), which for one stationary non-equilibrium state had the form $p(\gamma) = Z_{1\gamma} = \dfrac{1}{1 + \gamma \Gamma_{01}}$, $x = \gamma \Gamma_{01}$, now for an ensemble of stationary non-equilibrium states is equal to (59), $\varphi(\gamma) = Z_\gamma$,

$$Z_\gamma = \int e^{-\gamma \Gamma} p(\Gamma) \, d\Gamma = e^{-a \frac{x}{1+x}}. \tag{60}$$

Here, the subscript 1 denotes the quantities that describe one stationary non-equilibrium state. This also applies to changes in entropy: $\Delta_1$ is a change in the density of entropy in one stationary non-equilibrium state and $\Delta$ is change in entropy density in an ensemble of stationary non-equilibrium states. From (60), (6) and (31) we get

$$\ln Z_\gamma = -\frac{ax}{1+x}, \quad x = \gamma\Gamma_{01}, \quad \overline{\Gamma} = -\frac{\partial \ln Z_\gamma}{\partial \gamma} = \frac{a\Gamma_{01}}{(1+x)^2} + (-\ln Z_\gamma)\frac{1}{a}\frac{\partial a}{\partial \gamma}, \quad (61)$$

$$s = \beta\overline{u} + \gamma\overline{\Gamma} + \ln Z = \beta u_\beta + \ln Z_\beta - \Delta, \quad \overline{u} = u_\beta + u_\gamma.$$

Note that, as above, $u_\beta$ and $u_\gamma$ are average rather than random variables. Assume that the parameter $a$ depends on $\beta$ and $\gamma$. Then

$$u_\gamma = -\frac{\partial \ln Z_\gamma}{\partial \beta} = (-\ln Z_\gamma)\frac{1}{a}\frac{\partial a}{\partial \beta} + aZ_{1\gamma}u_{1\gamma} = \frac{x}{1+x}\frac{\partial a}{\partial \beta} + \frac{\partial x}{\partial \beta}\frac{a}{(1+x)^2}, \quad u_{1\gamma} = -\frac{\partial \ln Z_{1\gamma}}{\partial \beta}\bigg|_\gamma. \quad (62)$$

If for $\Gamma_{01}$ use expression (50), for $u_\gamma$ - expression (52), then $\frac{\partial \Gamma_{01}}{\partial \beta} = -\frac{\Gamma_{01}}{\beta}$, $\frac{\partial x}{\partial \beta} = -\frac{x}{\beta}$,

$$\beta u_\gamma + \gamma\overline{\Gamma} = (-\ln Z_\gamma)\frac{1}{a}m, \quad m = \beta\frac{\partial a}{\partial \beta} + \gamma\frac{\partial a}{\partial \gamma}, \quad \beta u_{1\gamma} + \gamma\overline{\Gamma}_1 = 0, \quad \overline{\Gamma}_1 = -\frac{\partial \ln Z_{1\gamma}}{\partial \gamma}\bigg|_\beta. \quad (63)$$

Substituting these expressions into the relation for the density of entropy (61), we obtain

$$-\ln Z_\gamma = a\Delta_2, \quad \Delta_2 = \frac{\Delta}{a-m}. \quad (64)$$

Substituting (64) into (61), (62) we obtain

$$1 - Z_{1\gamma} = \frac{(-\ln Z_\gamma)}{a} = \Delta_2, \quad \overline{\Gamma} = \Delta_2\frac{\partial a}{\partial \gamma} + a\frac{\Gamma_{01}}{(1+x)^2}. \quad (65)$$

Because $\Delta|_{\gamma=0} = 0$, then

$$\Gamma_0 = \Gamma|_{\gamma=0} = a\Gamma_{01}, \quad (66)$$

$$Z_\gamma = e^{-a(1-Z_{1\gamma})}, \quad Z_{1\gamma} = e^{-\Delta_1} \leq 1, \quad 1 - Z_{1\gamma} \geq 0, \quad 0 \leq Z_\gamma \leq 1, \quad a > 0, \quad Z_{1\gamma} = 1 - \Delta_2, \quad 0 \leq \Delta_2 \leq 1. \quad (67)$$

Growth condition $\overline{\Gamma}$, $\overline{\Gamma} > \Gamma_0$ has the form

$$\Delta_2\frac{\partial a}{\partial \gamma} + \frac{a\Gamma_{01}}{(1+x)^2} > a\Gamma_{01}. \quad (68)$$

Taking into account the additive energy when $\overline{u} = a\overline{u}_1$, from expression (62) for $u_\gamma$ we get

$$\beta\Delta_2\frac{\partial a}{\partial \beta} = -a(\frac{x}{1+x})^2 = -a(1-y)^2, \quad y = e^{-\Delta_1} = Z_{1\gamma}. \quad (69)$$

From expression (65)

$$\Delta_2 = 1 - e^{-\Delta_1} = \frac{\Delta}{a-m}.$$

Substituting this expression into (69), we obtain

$$\beta\frac{\partial a}{\partial \beta} = -a(1-y), \quad \Delta = (a-m)(1-y) = (1-y)[a + a(1-y) - \gamma\frac{\partial a}{\partial \gamma}],$$

$$\gamma\frac{\partial a}{\partial \gamma} = -\frac{\Delta}{1-y} + a + a(1-y), \quad m = a - \frac{\Delta}{1-y}. \quad (70)$$

The dependence of the number of stationary non-equilibrium states $a$ on the parameters $\gamma$ and $\beta$ can be illustrated, for example, by a change in the number of Benard cells depending on temperature and heat fluxes.

Substitution of (70) into (61) gives an expression for the average lifetime of the form

$$\overline{\Gamma} = a\Gamma_{01}[\frac{1}{(1+x)^2} + y(2-y-\frac{\Delta}{a(1-y)})] = \Gamma_{01}y\frac{[2a(1-y)-\Delta]}{(1-y)}. \quad (71)$$

For distribution with the Laplace transform (59), (60), the relationship of non-equilibrium energy and average lifetime takes the form (when $\Delta = a\Delta_1$)

$$\beta u_\gamma + \gamma \bar{\Gamma} = a(1-y) - \Delta = (-\ln Z_\gamma) - \Delta = a(1 - e^{-\Delta_1} - \Delta_1). \quad (72)$$

It is seen that the condition that the expression (72) is equal to zero is satisfied only in the linear approximation in a $\Delta_1$, when $1 - e^{-\Delta_1} \approx \Delta_1$. Substituting expressions (70) into inequality (68), we obtain the condition for the growth of the average lifetime $\bar{\Gamma}$, $\bar{\Gamma} > \Gamma_0$,

$$\frac{\Delta}{a} < \frac{(2y-1)(1-y)}{y}. \quad (73)$$

From (71), relation (73) is also written. If we assume local equilibrium in stationary non-equilibrium states, then we can use the additivity property of entropy, whence $\Delta = a\Delta_1$. Possible non-additive effects of entropy are considered small and neglected.

Then condition (73) is not satisfied for positive values $\Delta_1 > 0$. It can be shown that the growth condition for the average lifetime with the exponential distribution $Z_{1\gamma} = \frac{1}{1+\gamma\Gamma_{01}}$, $x = \gamma\Gamma_{01}$, is satisfied if instead of the complex Poisson distribution (60), the binomial distribution with $Z_\gamma = \int e^{-\gamma\Gamma} p(\Gamma) d\Gamma = [1 + p(Z_{1\gamma} - 1)]^n$ is used.

An increase in the average lifetime for one stationary non-equilibrium state can be obtained by using other distributions for the lifetimes in expressions (11)-(12) that are different from the limiting exponential distribution, for example, the Weibull, Gompertz distributions, etc.

So, for the Weibull distribution

$$f_X(x) = \frac{k}{\lambda}(\frac{x}{\lambda})^{k-1} e^{-(x/\lambda)^k}, \quad x \geq 0, \quad f_X(x) = 0, \quad x < 0, \quad (74)$$

where $k$ and $\lambda$ are distribution parameters, when the Laplace transform is written as a series

$$Z_{1\gamma} = \sum_{n=0}^{\infty} \frac{(-\gamma\lambda)^n}{n!} \Gamma(1+\frac{n}{k}) = 1 - \gamma\lambda\Gamma(1+\frac{1}{k}) + \frac{(\gamma\lambda)^2}{2!}\Gamma(1+\frac{2}{k}) - \frac{(\gamma\lambda)^3}{3!}\Gamma(1+\frac{3}{k}) + \dots, \quad (75)$$

where $\Gamma(\dots)$ is the gamma function, subject to the quadratic terms in $\gamma$, the condition $\bar{\Gamma}_1 > \Gamma_0$ is satisfied for

$$e^{\Delta_1}\sqrt{1 - 2n_2(1 - e^{-\Delta_1})}] > 1, \quad e^{\Delta_1} > 2n_2 - 1 \quad n_2 = \Gamma(1+2/k)/\Gamma(1+1/k). \quad (76)$$

When inequality (76) holds, the average lifetime increases in the case of one stationary non-equilibrium state. This is a significant difference between the Weibull distribution for the lifetime (74) and the exponential distribution (13).

For distribution

$$f(x) \sim e^{-px^2 - qx}, \quad (77)$$

which can be obtained using the maximum entropy method with constraints on the average value of $m$ and the second moment $m_2$, the Laplace transform is

$$Z_{1\gamma} = \frac{e^{(q+\gamma)^2/4p} erfc(\frac{q+\gamma}{2\sqrt{p}})}{e^{(q)^2/4p} erfc(\frac{q}{2\sqrt{p}})}, \quad (78)$$

where $erfc(z) = 1 - erf(z)$, $erf(z)$ is the probability integral. When $p \to 0$ from (78) we obtain that $Z_{1\gamma} \to 1/(1+\gamma/q)$. When $q = 1/\Gamma_{01p}$, where $\Gamma_{01p}$ is the average equilibrium lifetime of the exponential distribution, the function coincides with the exponential distribution function. The parameter $p$ shows the measure of deviation from the exponential distribution. Distribution (77) can lead to a normal distribution. In [27] probability densities for stationary solution of the Fokker Planck equation are probed in Gaussian approximation. However, for our purposes, the form (77) is more interesting, which clearly shows the degree of deviation from the limiting exponential distribution. From (78) we find

$$\overline{\Gamma}_1 = -\frac{\partial \ln Z_{1\gamma}}{\partial \gamma} = -\frac{(q+\gamma)}{2p} + Y_1, \quad Y_1 = \frac{e^{-z^2}}{\sqrt{\pi p}\, erfc(z)}, \quad z = \frac{(q+\gamma)}{2\sqrt{p}}, \quad \Gamma_0 = Y_{10} - \frac{q}{2p}, \quad Y_{10} = Y_1(z = z_0),$$

$$z_0 = \frac{q}{2\sqrt{p}}, \quad D_\Gamma = \langle \Gamma_1^2 \rangle - \langle \Gamma_1 \rangle^2 = \frac{\partial^2 \ln Z_{1\gamma}}{\partial \gamma^2} = \frac{1}{2p} - Y_1^2 + Y_1 \frac{(q+\gamma)}{2p}, \quad D_{\Gamma 0} = \frac{1}{2p} - Y_{10}^2 + Y_{10}\frac{q}{2p}.$$

Using the asymptotic expansion for $erfc(z) = 1 - erf(z)$, we obtain, in a linear approximation in (78), depending on $\gamma$ on $\Delta_1$, that the condition for the growth of the average lifetime for one stationary state, $\overline{\Gamma}_1 > \Gamma_{10}$, is satisfied for

$$y_{1-} < y_1 < y_{1+}, \quad y_{1+,-} = \frac{1}{12(1-a_1)^2} \frac{[11 - 47a_1 + 45a_1^2 - 9a_1^3]}{(1 - 3a_1)}[1 \pm \sqrt{D}],$$

$$\sqrt{D} = \frac{\sqrt{1 - 26a_1 + 247a_1^2 - 1068a_1^3 + 2223a_1^4 - 2106a_1^5 + 729a_1^6}}{(11 - 47a_1 + 45a_1^2 - 9a_1^3)},$$

$$y_1 = e^{-\Delta_1}, \quad a_1 = 2p/q^2 = \frac{2c_1(1-3c_1)}{(1+3c_1)^2}, \quad p \approx \frac{c_1}{4\Gamma_0^2}(1-3c_1), \quad c_1 = \frac{\Gamma_0^2}{\langle \Gamma_0^2 \rangle}, \quad q = \frac{1}{2\Gamma_0}(1+3c_1)$$

The growth conditions for the average lifetime are also satisfied when substituting the Laplace transform as an argument in the generating functions that describe the behavior of the ensemble of stationary non-equilibrium states. You can specify many such examples, specify many distribution functions for the lifetime, the use of which leads to an increase in the average lifetime.

The non-equilibrium part of the partition function $Z_\gamma$ reduces to the Laplace transform with respect to the non-equilibrium parameter $\gamma$ of the distribution of the lifetime, the first-passage time the random process first reaches the zero level. The results depend on the studied physical processes and the corresponding random process for energy, the distribution of the exit time for this random process. Of course, there is a wide variety of results. So, for example, for the fractional Brownian motion there is no increase of the average lifetime, but for the Feller process there is increase. In [26], a Gaussian approximation is used in the problem of transition through a potential barrier. For other processes, other expressions are possible for the probability density of the first passage of a level. In [56], for the probability density of the first passage of the level in the case of Brownian motion, an inverse Gaussian distribution density was obtained depending on the boundary conditions and the form of the time dependence of the drift and diffusion coefficients. In [57], a generalization of the Eyring – Kramers transition rate formula to irreversible diffusion processes was obtained. In [58] the Weibull distribution is obtained for the probability density of the first level passage. In [59], the escape probability is obtained for the Feller process. In [32] the limit exponential distribution for this quantity is received. In [60] provided an exact unified framework for studying the full statistics of first passage time under detailed balance conditions. In [61] determined exactly the first passage time distribution from the corresponding relaxation spectrum. In [62] explained the behavior of smooth first-passage time densities, for which all moments are finite. Recent reviews in this area [63 – 64].

## 5. Conclusion

We managed to explore a generalizing physical thermodynamic value – the lifetime of complex dynamical systems. The thermodynamic way of description aims at elucidating the general properties and statistical laws, which do not depend on the peculiarities of the matter structure and are universal invariants. The contents of the thermodynamics are the set of such results. Since the finiteness of lifetime of arbitrary realistic systems is a universal property, we believe its inclusion into the thermodynamic description to be fruitful and necessary.

The regularities established in the work make it possible to find the lifetime of physical, chemical, and other systems, and to model their behavior, exploring different features of their evolution and stationary states, taking into account different features of the past of systems. Thermodynamics with a lifetime describes open systems far from equilibrium and can be used to study the behavior of dissipative structures and other synergetic effects. It is possible to describe the non-equilibrium behavior of arbitrary physical quantities by which the investigated physical system is open, taking into account the influence of all factors that contribute to the interaction of the system with the environment.

This paper uses a general thermodynamic approach in which the details of stochastic distributions are not specified, although the results depend on them significantly. Some "basic" distributions for the lifetime are considered. The explicit form of these distributions and their parameters depend on the specific physical problem. The type of distributions for the lifetime is considered in a number of works [17, 23, 25-34, 56-67] and many other books and articles. In [65] studied the statistics of infima, stopping times and passage probabilities of entropy production in non-equilibrium steady states. In [65], expressions are obtained for the passage probabilities of entropy production through entropy production. To find the average lifetime, you need to know the explicit time dependencies for entropy production. In [66] developed a thermodynamic theory for processes in mesoscopic systems that terminate at stopping times, which generalize first-passage times. Stopping times in [66] are considered for various quantities, for example the escape of a particle from a metastable state. An analog of expression (2) of this paper in [66] is used (Eq. (14) from [66]), averaging over the probability distribution of the stopping time $T$.

However, the approach proposed in this paper allows us to relate the known type of distribution for the lifetime and changes in the lifetime, for example, the possibility of increasing the average lifetime with a change in entropy, fluxes and other thermodynamic characteristics of the statistical system. Using the approaches of this article, we can calculate the possibility of increasing or decreasing the lifetime of stationary nonequilibrium states under various kinds of influences on the system. The correspondence between the behavior of physical systems and stochastic models was studied in a number of works (for example, in [67] for harmonically trapped particle whose motion is described by a one – or multidimensional Ornstein – Uhlenbeck process).

The introduction of thermodynamic forces $\gamma$ conjugated to the value of $\Gamma$, corresponds to the effective consideration of the influence of external influences (such as a pump parameter, etc.). The introduction of the quantities $\gamma$ and $\Gamma$ corresponds to an effective account of the openness of the system.

In addition to the macroscopic thermodynamic approach, the methods used are statistical physics and stochastic theory, which are necessary in the description of open systems. The probabilistic task of the distribution of the lifetime contains a large number of different possibilities. These include various methods of the stochastic theory of storage and queuing theory.

Explicit expressions are obtained for the Laplace transform of the distribution of the system lifetime (first-passage time). Using this Laplace transform, the moments of a random value of the lifetime are recorded, first, the average values of the lifetime for non-equilibrium state. Explicit relations are also obtained for the conjugate of lifetime of the thermodynamic variable $\gamma$, which is expressed through changes in the entropy of the system. In the first approximation of the expansion of the exponential expression, the quantity $\gamma$ is equal to the production of entropy, which coincides with the approximation obtained from the *NSO* method (expression (2)). The characteristics of the lifetime in non-equilibrium state are expressed in terms of the conjugate lifetime parameter $\gamma$ and, accordingly, through the entropy characteristics of the system. The dependences of the flux $q$ on the non-equilibrium parameter $\gamma$, a number of relations for the non-equilibrium temperature $\theta^{-1}$ from [2], for the flux $q$, non-equilibrium energy $u_\gamma$ and average lifetime $<\Gamma>$ are obtained.

In this work, we use mainly the limiting exponential distribution, which is practically not realized (although in [23] its proximity to distributions describing the time to reach the level was estimated). For other distributions for the lifetime (for example, from the queuing theory), which are valid for small time intervals, it is possible to increase the average lifetime of the system with a positive increase in the entropy of the system.

The non-equilibrium part of the partition function is highlighted $Z_\gamma$ from $Z(\beta,\gamma) = Z_\beta Z_\gamma$, where $Z_\gamma = \int e^{-\gamma\Gamma} p(\Gamma)\,d\Gamma$, $p(\Gamma)$ is lifetime distribution, and non-equilibrium free energy. Expression $<u_\gamma> = -\partial \ln Z_\gamma / \partial \beta |_\gamma$ gives non-equilibrium internal energy equal to zero in equilibrium; $\bar{u} = -\dfrac{\partial \ln Z}{\partial \beta} = u_\beta + u_\gamma$, $u_\beta = -\dfrac{\partial \ln Z_\beta}{\partial \beta}\bigg|_\gamma$. The functional for a random value of the lifetime, in terms of stochastic theory - a process subordinate to the main random process - energy, depends on energy, and the introduction of a thermodynamic variable of the lifetime corresponds to some redundancy of the description (albeit additional information), but makes it possible to explicitly express the lifetime through changes in the entropy of the system, which is what is done in the present work. If we write the distribution (4) in the form $\rho \sim e^{-L}$, $L = \beta u + \gamma \Gamma$, where $u$ and $\Gamma$ are random variables, then for one stationary non-equilibrium state, exponential distribution (13) for the lifetime and expressions (39), (45), (50), (63) we obtain $\beta u_\gamma + \gamma \bar{\Gamma} = 0$, $\bar{L} = \beta u_\beta$, and averaging the value of $L$ leads to the equilibrium case. Thus, the introduction of the thermodynamic variable of the lifetime and the redundancy of the description in this situation is manifested at the level of fluctuations. But the partition function changes, the normalization of the form $\int e^{-L} \omega(u,\Gamma)\,du\,d\Gamma$, $L = \beta u + \gamma \Gamma$, where random variables appear in $L$. For a complex Poisson distribution, an ensemble of stationary non-equilibrium states, as well as other distributions for the lifetime, the situation is different.

Random values also differ $u$, $\Gamma$ from distribution (4) and average values $<u>$, $<\Gamma>$, $u_\beta$, $u_\gamma$, which appear in thermodynamic relations.

Obvious dependences of the Laplace transform of the distribution of the lifetime, average lifetime, and the conjugated thermodynamic quantity $\gamma$ on the change in entropy upon the transition of the system to another stationary non-equilibrium state are obtained. The dependences for the derivatives on the average number of stationary non-equilibrium states $a$ with respect to the reciprocal temperature $\beta$ and non-equilibrium parameter $\gamma$ are also obtained. It is noted that the behavior of one stationary non-equilibrium state differs from the behavior of a collective of stationary non-equilibrium states.

It is shown that in the case of entropy additivity and a positive deviation of the system entropy from the equilibrium value, the choice of the limiting exponential distribution for describing the lifetime does not make it possible to describe the growth of the average lifetime. This is due to the limit behavior of the exponential distribution, which is valid in the limit of infinitely large times. However, in these cases, lifetimes at finite time intervals are considered. A description of the growth of the average lifetime for one stationary non-equilibrium state is possible when choosing distributions for the lifetime, differing from the limiting exponential distribution.